\documentclass[sigconf,9pt,nonacm]{acmart}

\usepackage{amsmath,amssymb,amsfonts}
\usepackage{graphicx}
\usepackage{textcomp}
\usepackage{amsmath}
\usepackage{booktabs} 
\usepackage{multirow}
\usepackage[most]{tcolorbox}      
\tcbuselibrary{listings,skins,breakable}
\usepackage{listings}  
\usepackage{hyperref}       
\usepackage{url}  
\usepackage{setspace}
\usepackage{booktabs}       
\usepackage{amsfonts}       
\usepackage{nicefrac}       
\usepackage{microtype}      
\usepackage{xcolor}         
\usepackage{algorithm}
\usepackage{algpseudocode}
\usepackage{amsmath}
\usepackage{tikz}

\usepackage{orcidlink}
\usepackage{comment}
\usepackage{amssymb}
\usepackage{braket}
\usepackage{xcolor}
\usepackage[font=scriptsize,skip=0pt]{caption}
\usepackage{cleveref}
\usepackage{setspace}
\def\BibTeX{{\rm B\kern-.05em{\sc i\kern-.025em b}\kern-.08em
    T\kern-.1667em\lower.7ex\hbox{E}\kern-.125emX}}

\usepackage{enumitem}
\usepackage{hyperref}
\setlist[itemize]{leftmargin=*}%
\setlist[enumerate]{leftmargin=*}%

\usepackage[most]{tcolorbox}
\usepackage{hyperref}
\usepackage{cleveref}

\newtcolorbox[auto counter, number within=subsection]{mybox}[2][]{%
  enhanced,
  colback=black!3!white,
  colframe=black!90!black,
  breakable,
  left=0.5pt, right=0.5pt, top=0.5pt, bottom=0.5pt,
  boxsep=0.5pt, boxrule=0.5pt,
  title=Box~\thetcbcounter\ -- #2,
  label type=mybox,
  #1
}

\crefname{mybox}{Box}{Boxes}
\Crefname{mybox}{Box}{Boxes}

\usepackage{fancyhdr}
\fancypagestyle{firstpage}{
   \fancyhf{} 
   \fancyhead[C]{\textcopyright\ 2026 IEEE/ACM. This is the author’s version of the work. The definitive Version of Record will be Published in IEEE/ACM Design Automation Conference (DAC) Proceedings, Long Beach, CA, United States, July 2026.} 
}

\begin{document}
\settopmatter{printfolios=false}


\title{Closing the Loop: Resource-aware Hybrid NAS Guided by Analytical and Hardware-Calibrated Quantum Cost Modeling\\
\vspace{-10pt}
}





\author{Muhammad Kashif, Alberto Marchisio, Muhammad Shafique}
\affiliation{%
  \institution{eBRAIN Lab, Division of Engineering, New York University Abu Dhabi}
  \streetaddress{PO Box 129188}
  \city{Abu Dhabi}
  \country{United Arab Emirates}
}

\affiliation{%
  \institution{Center for Quantum and Topological Systems, NYUAD Research Institute, New York University Abu Dhabi}
  \city{Abu Dhabi}
  \country{United Arab Emirates}
}
\email{{muhammadkashif, alberto.marchisio, muhammad.shafique}@nyu.edu}

\begin{abstract}
\begin{spacing}{0.94}
Hybrid quantum–classical neural networks (HQNNs) integrate quantum circuits with classical layers, each operating under fundamentally different computational paradigm, which makes hardware-resource estimation challenging. Also, the training of quantum circuits on real devices requires thousands of circuit executions, which is impractical on current NISQ devices. Therefore, most HQNNs are evaluated on classical simulators, with hardware cost approximated using floating-point operations (FLOPs). However, FLOPs and existing quantum resource estimation methods (e.g., gate counts) overlook key quantum hardware-specific factors such as gate durations, limited qubit connectivity, and noise, all of which ultimately determine the true cost and scalability of quantum circuits.
%

In this paper, we propose an analytical quantum cost model that estimates quantum hardware resources using real backend calibration data, incorporating gate durations, routing overheads, and noise-induced sampling inefficiencies. To complement this, we develop a classical cost model that converts FLOPs into device-specific throughput, enabling a unified time-based representation of hardware-resource cost for both subsystems of HQNNs. Building on these analytical models, we present Hyb-HANAS, a hardware-aware hybrid neural architecture search framework, which jointly optimizes accuracy, hardware cost, and parameter count using NSGA-II. Hyb-HANAS identifies Pareto-optimal trade-offs and cross-domain co-adaptation between classical and quantum components of HQNNs. Beyond NAS, the proposed quantum analytical cost model is broadly applicable to quantum hardware benchmarking, compiler evaluation, and training-time estimation of quantum circuits on NISQ devices.
\end{spacing}
\vspace{-1pt}
\end{abstract}

\maketitle

\thispagestyle{firstpage}

\begin{spacing}{0.94}

\vspace{-8pt}
\section{Introduction}
\vspace{-2pt}

In classical neural networks (NNs), hardware resource-aware neural architecture search (HANAS), frequently uses floating-point operations (FLOPs) to estimate computational resources\cite{wu2019fbnet,chitty2022neural}. FLOPs provide a hardware-agnostic proxy for the arithmetic complexity, making them a convenient surrogate for runtime and energy consumption across different devices \cite{roshtkhari:2025}. By penalizing architectures with excessive FLOPs, HANAS frameworks identify Pareto-optimal models that balance accuracy and computational cost.
While effective for classical NNs and recently extended to HQNNs \cite{kashif:2025_DAC,kashif2025faqnas}, these FLOPs-based metrics are fundamentally inadequate for hybrid quantum-classical neural networks (HQNNs), where computations span both classical and quantum layers (Fig.~\ref{fig:F_and_t_vs_Q_andD}-top). Quantum operations are not tensor computations, and their hardware cost depends on quantum-specific factors, such as qubit connectivity, noise, and decoherence. Quantum circuits are executed on dedicated qubit simulators, which are invisible to standard FLOPs profiling tools. Consequently, FLOPs-based profiling treats quantum layers as a \textit{black box} and significantly underestimates the true computational footprint of HQNNs.
This disconnect breaks the feedback loop between architecture design and actual hardware cost. Closing this loop requires embedding a hardware-calibrated quantum cost model within the hardware cost-estimation process, linking high-level architectural designs with the physical constraints of quantum devices.

\textbf{State-of-the-art} on quantum hardware resource estimation is largely underexplored. The closest is the recent ML-based quantum runtime predictor~\cite{ma:2024}, which trains an ML model on $\approx$1500 quantum circuit execution times to predict new circuit runtimes. However, this is effectively a learned surrogate for Qiskit’s scheduler \cite{qiskit_scheduler}, as the runtime of a quantum circuit is mainly governed by physical gate-execution windows. These methods only model raw gate timing, ignoring the statistical impact of gate errors and noise, which mainly increases the unreliability in results (more details in Section \ref{sec:Qcost_model}). 
Additionally, from QML perspective, factors such as gradient evaluation overhead are also ignored in these models. \textit{Our analytical cost model combines calibrated gate-level timing with noise-induced sampling overhead and gradient costs (for QML algorithms), providing a realistic estimate of training latency rather than just circuit runtime.}

From NAS perspective, existing work focuses mainly on accuracy optimization, occasionally including abstract quantum hardware metrics like gate count and circuit depth \cite{zhang2022,du2022quantum,li2023eqnas,li2024balanced,dutta2025}. These methods typically target ideal simulators without physical device calibration data and fail to combine quantum and classical hardware costs, which is critical for HQNNs or hybrid models in general. \textit{In this paper, we integrate a hardware-calibrated quantum cost model with an aligned classical cost model into a unified time-based cost metric for multi-objective hybrid NAS.}

%
%
\vspace{-8pt}
\subsection{Motivational Analysis} \label{subsec:mot_analysis}
\vspace{-3pt}
%
%

\begin{figure}[t!]
    \centering
    \vspace{-7pt}
    \includegraphics[width=1.0\linewidth]{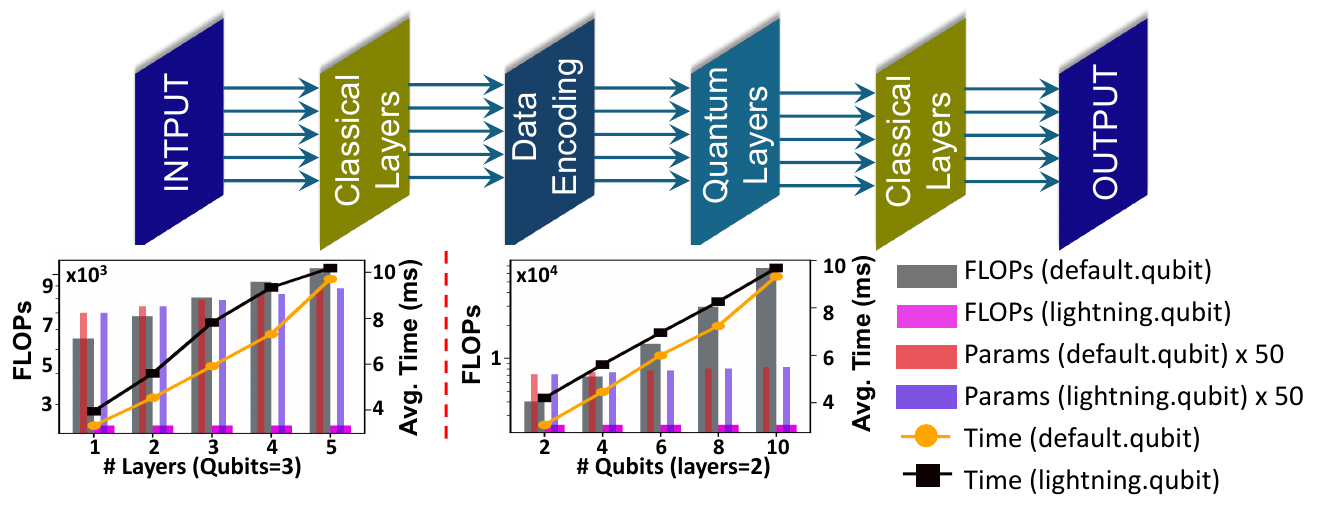}
    \vspace{-12pt}
    \caption{(Top) An abstract illustration of HQNN. (bottom) FLOPs, parameter count, and wall-clock time (averaged over 5 runs) comparison of quantum circuits for different depths and qubit counts on two different backend quantum simulators. Since the FLOPs are in the order of thousands compared to the parameters in the order of hundreds, the parameter count is scaled by 50 for better visualization.}
    \label{fig:F_and_t_vs_Q_andD}
    \vspace{-18pt}
\end{figure}

To demonstrate FLOPs limitation in HQNNs, we implemented a simple HQNN with fixed classical layers interleaved with quantum layers (angle embedding \cite{kashif2021design} + strongly entangling layer \cite{SEL}) in PennyLane \cite{qlayers_pytorch}. 
We measured FLOPs for two different qubit simulators:
(1) \texttt{default.qubit}, a Python-based state-vector backend~\cite{default_qubit}, and
(2) \texttt{lightning.qubit}, an optimized C++/CUDA simulator~\cite{lightning_qubit}. 
With \texttt{default.qubit}, the measured FLOPs consistently increase with circuit depth and qubit count (Fig.~\ref{fig:F_and_t_vs_Q_andD}-bottom), as quantum simulation 
is implemented as PyTorch tensor operations, which are visible to the FLOPs profiler. However, this effectively makes the FLOPs reflect the classical simulation overhead rather than the true quantum cost, as used in~\cite{kashif:2025_DAC,kashif2025faqnas}.
In contrast, \texttt{lightning.qubit} performs its computations inside optimized external kernels that are external to PyTorch’s computation graph, resulting in constant FLOPs across different circuit sizes, even though the actual quantum simulation cost increases, and therefore the run time increases, despite constant FLOPs, indicating that FLOPs alone fail to reflect the actual computational cost of HQNNs.
This discrepancy is even more pronounced on real hardware, as quantum circuits are completely offloaded to the hardware backend, and contribute zero FLOPs.
Although the parameter count grows with circuit size, it can also be misleading, because standard parameter counters 
only count single-qubit rotation gates as parameters and ignore entangling gates, which dominate both expressibility and hardware cost. Hence, parameter count is also insufficient as a proxy for the computational cost of quantum layers in HQNNs.

\begin{figure}[t!]
    \centering
    \vspace{-8pt}
    \includegraphics[width=1.0\linewidth]{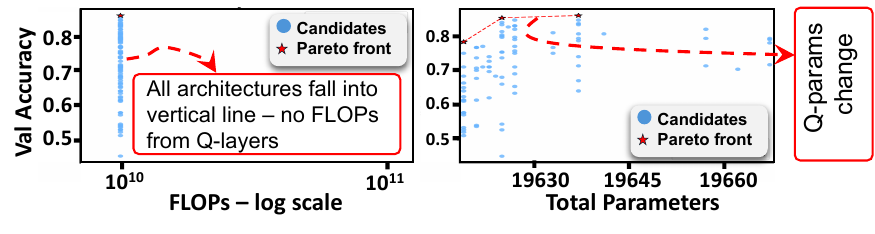}
    \vspace{-15pt}
    \caption{FLOPs vs. accuracy (left) and parameters vs. accuracy (right) plots for hybrid NAS with fixed classical components. Since quantum layers contribute no FLOPs, the FLOPs-accuracy plot collapses into a vertical line, producing an uninformative Pareto front. In contrast, parameter-based profiling captures quantum layer variations, yielding a more meaningful distribution of architectures and a relatively well-defined Pareto front.}
    \label{fig:mot_study_NAS}
    \vspace{-18pt}
\end{figure}

Similarly, from HANAS perspective, when classical layers are fixed and only the quantum layer varies, FLOPs vs. accuracy collapses into a vertical line (Fig.~\ref{fig:mot_study_NAS}, left), with an uninformative Pareto front, mainly because quantum layers add no FLOPs. 
On the other hand, standard profiling tools count quantum layer’s trainable parameters in total count, primarily from single-qubit gates, thus resulting in a more distributed set of architectures and a relatively meaningful Pareto front (Fig.~\ref{fig:mot_study_NAS}, right). 

\textit{In summary, FLOPs-based hardware-resource estimation for quantum layers, at best, serve as a proxy for classical simulation overhead (in certain qubit simulators), not for the true computational footprint on quantum hardware.
In the context of HANAS for HQNNs, relying solely on FLOPs can lead to misleading conclusions, as architectures may appear efficient in terms of FLOPs but are impractical on real quantum devices. 
To make HANAS meaningful for HQNNs, the hardware cost model must reflect both classical and quantum components of the hybrid stack. 
This requires a unified cost metric that integrates quantum hardware–aware costs for quantum layers with classical cost metrics for classical layers, providing a consistent and realistic measure of the total computational workload. Such a unified metric enables HANAS to properly evaluate hardware feasibility and guide the search toward architectures that are not only accurate but also feasible on actual quantum devices.}

\vspace{-8pt}
\subsection{Our Contributions}
\vspace{-3pt}

\begin{itemize}
    \item \textbf{Identification of FLOPs Limitations in Hybrid Models.} We identify and demonstrate that conventional FLOPs-based metrics fail to capture the true computational cost of HQNNs (\textbf{Sec.~\ref{subsec:mot_analysis}}). 

    \item \textbf{Analytical quantum cost modeling.}
    It estimates quantum hardware resources (in terms of time) directly from real backend calibration data in real-time. Our proposed model integrates calibrated gate durations, transpilation-induced routing overhead, and reliability factors such as gate and decoherence errors into a single effective time-based cost metric, enabling realistic, hardware-specific resource estimation (\textbf{Sec.~\ref{sec:Qcost_model}}).

    \item \textbf{Analytical classical cost modeling.}
    To complement the proposed analytical quantum cost model, we develop a classical cost formulation based on exact FLOPs count, which are then converted into equivalent time units using device-calibrated throughput, allowing classical and quantum components to be expressed on a common temporal scale. This unified cost representation enables direct and fair comparison across different hybrid architectures (\textbf{Sec.~\ref{sec:classical_cost_model}}).

    \item \textbf{Hardware resource-aware hybrid neural architecture search (Hyb-HANAS).}
    The analytical cost models are then integrated as a unified cost into the multi-objective NAS. We then propose the Hyb-HANAS framework that jointly optimizes accuracy, classical cost, quantum cost, and total parameters. Each candidate hybrid network is evaluated using both accuracy and backend-calibrated hardware-resource estimation, achieving truly hardware-informed Pareto optimization (\textbf{Sec.~\ref{sec:HANAS_framework}}).

    \item \textbf{Insights for Hardware-Aware HQNN Design.}
    Systematic analysis across varying circuit widths, depths, and entanglement topologies reveals how routing and noise-induced overhead influence the resource-performance trade-offs in HQNNs. 
    Our analysis also identifies the components that dominate quantum hardware cost, emphasizing the importance of hardware-calibrated cost models and motivating quantum-aware metrics for future NAS and benchmarking studies (\textbf{Sec.~\ref{sec:results}},~\textbf{\ref{sec:conclusion})}.

\end{itemize}

\vspace{-10pt}
\section{ Hardware-Aware Time-Based Cost Modeling}
\vspace{-4pt}
\subsection{Need for a Unified Cross-Domain Metric}
\vspace{-2pt}

Classical and quantum hardware are fundamentally different. Hence, estimating the hardware cost in hybrid algorithms, e.g., HQNNs, is typically challenging. The hardware cost for classical devices can be determined by arithmetic workload (FLOPs/sec), whereas quantum hardware cost is usually governed by calibrated gate durations, routing delays, and noise-induced sampling inefficiencies. Despite this mismatch, both types of cost naturally reduce to a common physical quantity, i.e., time. By converting classical FLOPs into device-specific latency via measured throughput and by computing quantum hardware latency from hardware calibration data, we express both subsystems in a unified time-based metric. 
Moreover, using time as a unified metric is intuitive and practical, as it directly reflects real hardware constraints 
(latency, power consumption, and throughput). Unlike abstract workload measures, a time-based formulation inherently captures device-dependent factors, such as memory access and tensor-core utilization for classical components, and qubit connectivity and decoherence effects for quantum circuits. This 
ensures that HQNN optimization targets true hardware efficiency rather than abstract device-agnostic computational complexity.



\vspace{-8pt}
\subsection{Classical Cost Modeling} \label{sec:classical_cost_model}
\vspace{-2pt}

\begin{figure*}[h]
    \centering
    \includegraphics[width=1.0\linewidth]{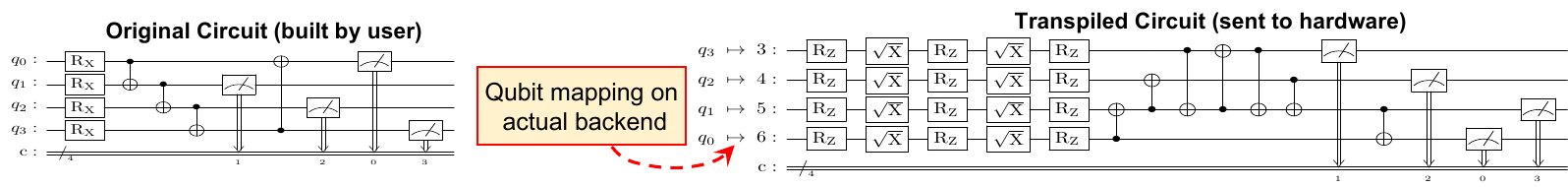}
    \caption{Original vs. Transpiled circuits. For a small $4$-qubit circuit, the single- and two-gate count, respectively, increased from $8$ to $24$ and $4$ to $7$, post transpilation. Transpilation penalizes circuits with more qubits by inserting more SWAP gates due to limited qubit connectivity in NISQ devices, i.e., more active qubits thus lead to higher routing complexity.}
    \label{fig:org_trans_ckt}
    \vspace{-14pt}
\end{figure*}

Classical training cost is estimated via throughput-based calibration from a reference model, as outlined in Algorithm~\ref{alg:classical_cost}.

%


\noindent The throughput $\Phi_{\text{device}}$ is computed using FLOPs for a given reference model. 
While the total time could be estimated directly by timing a full training iteration, the FLOPs-based throughput calibration offers several advantages. 
(1)~\textit{algorithmic complexity} (measured via FLOPs) is decoupled from \textit{hardware efficiency} (measured via throughput), enabling consistent comparison across different devices, as two models with the same FLOPs may exhibit different execution times on different hardware backends.
(2)~It provides stable and reproducible estimates, as FLOPs counts are deterministic and unaffected by transient runtime noise. 
(3)~This formulation aligns with the quantum cost model (details in Sec.~\ref{sec:Qcost_model}), where the number of quantum gates and error rates are converted to physical time. Consequently, both classical and quantum costs are expressed in comparable physical units (seconds), enabling balanced multi-objective optimization in the hybrid NAS.

\vspace{-6pt}
\begin{algorithm}[H]
\footnotesize
\caption{Classical Cost Model 
}
\label{alg:classical_cost}
\begin{algorithmic}[1]
\Require Reference model FLOPs $F_{\text{model}}$ and measured time $T_{\text{measured}}$
\Require Candidate model FLOPs count, training steps $N_{\text{steps}}$

\State \textbf{\footnotesize Step 1: Calibrate device throughput from reference model}
\State $\Phi_{\text{device}} \gets \frac{F_{\text{model}}}{T_{\text{measured}}}$ \Comment{\footnotesize FLOPs/sec}

\State \textbf{\footnotesize Step 2: Estimate per-step time for any candidate model}
\State $T_{\text{classical}} \gets \frac{F_{\text{model}}}{\Phi_{\text{device}}}$ \Comment{\footnotesize Time from FLOPs count}

\State \textbf{\footnotesize Step 3: Scale to total training duration}
\State $T_{\text{classical,total}} \gets T_{\text{classical}} \times N_{\text{steps}}$

\State \Return $T_{\text{classical,total}}$
\end{algorithmic}
\end{algorithm}
\vspace{-13pt}

\vspace{-9pt}
\subsection{From Logical to Physical Circuits}
\vspace{-2pt}

Before going into the details of the proposed analytical quantum cost model, it is essential to distinguish between logical and physical circuits. \mbox{Logical} circuits, defined in frameworks like PennyLane or Qiskit, assume ideal connectivity and universal gate access, whereas real quantum hardware imposes constraints such as limited qubit connectivity and native gate sets. During \emph{transpilation} (an intermediary step that maps a logical circuit to hardware), non-native gates are decomposed into device-native gates and additional operations (e.g., SWAPs) are inserted to satisfy hardware connectivity, often increasing circuit depth.
As illustrated in Fig.~\ref{fig:org_trans_ckt}, transpiling a simple 4-qubit logical circuit to the \texttt{ibm\_brisbane} backend substantially increases its depth and gate count. Therefore, realistic hardware cost estimation must be based on the transpiled circuit rather than the logical design.



\vspace{-8pt}
\subsection{Analytical Quantum Cost Model} \label{sec:Qcost_model}
\vspace{-2pt}

We model quantum hardware resource cost by combining physical gate times, routing overhead, noise-induced sampling inefficiency, and gradient evaluation requirements. The complete 
model is presented in Algorithm~\ref{alg:qcost_model}, with some important considerations discussed below.

\noindent\textit{\textbf{Routing overhead.}} Limited qubit connectivity in NISQ devices requires additional two-qubit gates to route non-adjacent qubit interactions. Since $t_{2q} \approx 200$--$350\,\mathrm{ns}$ versus $t_{1q} \approx 30$--$50\,\mathrm{ns}$, routing cost is usually dominated by two-qubit operations.

\noindent\textit{\textbf{Noise-induced reliability overhead.}} The factor $(1-p_{\text{fail}})^{-1}$ in Steps 3 and 4 of Algorithm~\ref{alg:qcost_model} represents the statistical overhead modeled as an \emph{effective time penalty} to account for hardware reliability due to qubit decoherence and gate errors. Note that noise does not increase the physical runtime of a circuit, and quantum processors execute each shot in the same fixed duration (depending on gate durations), regardless of whether the outcome is reliable or corrupted. Instead, noise reduces the fraction of useful measurement results, which effectively means that more shots are statistically required to obtain one accurate sample. Thus, this factor is not a literal hardware retry but an effective time penalty that accounts for noise-induced sampling inefficiency. This is similar to requiring more transmissions in a noisy communication channel, for instance, with a $90\%$ success probability, approximately $1/0.9 \approx 1.11$ executions are needed per valid sample. 

\noindent\textit{\textbf{Gradient evaluation.}} Real quantum devices use the parameter-shift rule for gradient evaluation~\cite{Schuld_2019,Mitarai_2018}, requiring two circuit evaluations per parameter at $\theta_i \pm \pi/2$, unlike simulators which also support single-step gradient estimation, such as adjoint differentiation \cite{adjoint_diff}. 
The total hardware cost (in seconds), for a given hybrid model, simply combines classical and quantum costs (outputs of Algorithms~\ref{alg:classical_cost} and~\ref{alg:qcost_model}):
\vspace{-4pt}
\begin{equation}
\vspace{-4pt}
C_{\text{total}} = T_{\text{classical, total}} +  T_{\text{quantum, total}}
\label{eq:total_cost}
\end{equation}
By expressing both classical and quantum components' cost in time units, this unified cost model provides a deterministic, \mbox{reproducible}, and physically interpretable measure of a hybrid network's hardware resource requirements, which is device-agnostic in structure yet calibrated to specific hardware for fair and realistic comparison across different quantum backends.
\begin{figure}[t!]
\vspace{-6pt}
\begin{algorithm}[H]
\footnotesize
\caption{Analytical Quantum Cost Model}
\label{alg:qcost_model}
\begin{algorithmic}[1]
\Require Transpiled circuit with $N_{1q}$, $N_{2q}$, $N_{\text{meas}}$ operations  \Comment{\footnotesize  \# of 1qubit, 2qubit and measurement gates}
\Require Backend calibration: $t_{1q}$, $t_{2q}$, $t_{\text{meas}}$, $\epsilon_{1q}$, $\epsilon_{2q}$, $\epsilon_{\text{meas}}$, $T_2$ \Comment{\footnotesize gate durations and error rates of 1qubit, 2qubit and measurement gates, $T_2$ is qubit's coherence time}
\Require Training params and total steps: $N_{\text{params}}$, $N_{\text{steps}}$

\State \textbf{\footnotesize Step 1: Compute combined gate execution time}
\State $T_{\text{gate}} \gets N_{1q} \cdot t_{1q} + N_{2q} \cdot t_{2q} + N_{\text{meas}} \cdot t_{\text{meas}}$ \Comment{\footnotesize Total gate execution time (transpiled)}

\State \textbf{\footnotesize Step 2: Quantify routing overhead from excess native gates}
\State $T_{\text{routing}} \gets \sum_{g \in \mathcal{G}_{2q}} \Delta N_{2q}^{(g)} \cdot t_{2q}^{(g)}$  \Comment{\footnotesize Routing overhead}
\State \hspace{1em} where $\Delta N_{2q}^{(g)} = N_{2q,\text{phys}}^{(g)} - N_{2q,\text{logical}}^{(g)}$ 
\State $T_{\text{logical}} \gets T_{\text{gate}} - T_{\text{routing}}$ \Comment{\footnotesize Logical circuit execution time}

\State \textbf{\footnotesize Step 3: Reliability - Modeling circuit failure probability}
\State $p_{\text{gate}} \gets 1 - (1-\epsilon_{1q})^{N_{1q}} (1-\epsilon_{2q})^{N_{2q}} (1-\epsilon_{\text{meas}})^{N_{\text{meas}}}$ \Comment{\footnotesize Cumulative gate errors}
\State $p_{\text{decoh}} \gets 1 - e^{-T_{\text{gate}}/T_2}$ \Comment{\footnotesize Decoherence loss probability}
\State $p_{\text{fail}} \gets 1 - (1-p_{\text{gate}})(1-p_{\text{decoh}})$ \Comment{\footnotesize Combined failure probability}

\State \textbf{\footnotesize Step 4: Adjust for sampling inefficiency}
\State $T_{\text{eff}} \gets \frac{T_{\text{logical}} + T_{\text{routing}}}{1 - p_{\text{fail}}}$ \Comment{\footnotesize  Noise-corrected time}

\State \textbf{\footnotesize Step 5: Account for gradient evaluation using parameter-shift}
\State $N_{\text{eval}} \gets 2 \cdot N_{\text{params}}$ \Comment{\footnotesize Two evals per parameter}
\State $T_{\text{quantum}} \gets T_{\text{eff}} \times N_{\text{eval}}$ \Comment{\footnotesize Per-step cost}

\State \textbf{\footnotesize Step 6: Scale to full training duration}
\State $T_{\text{quantum,total}} \gets T_{\text{quantum}} \times N_{\text{steps}}$ \Comment{\footnotesize Total training time}

\State \Return $T_{\text{quantum,total}}$
\end{algorithmic}
\end{algorithm}
\vspace{-20pt}
\end{figure}
%
\begin{figure*}
    \includegraphics[width=\linewidth]{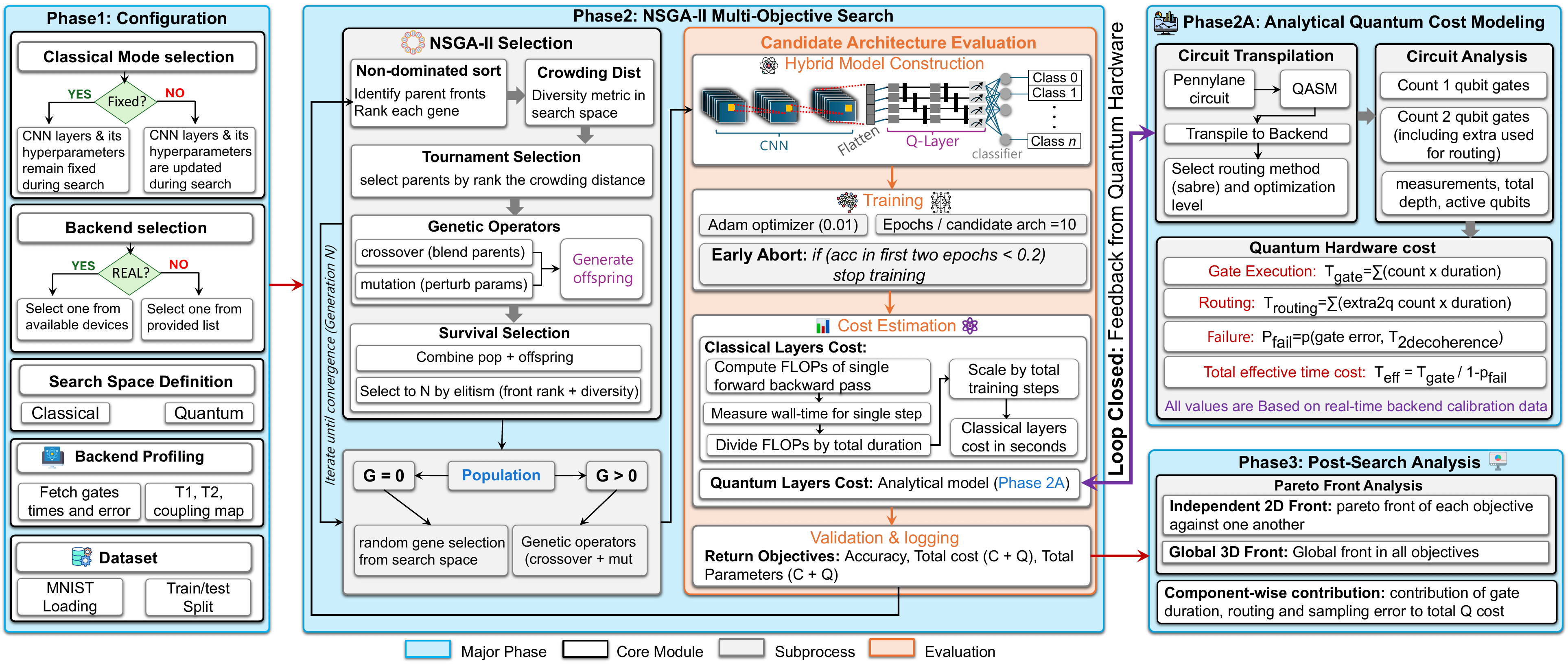}
    \caption{Overview of the Hyb-HANAS framework. The workflow includes configuration setup, NSGA-II-based multi-objective evolutionary search, and hardware-aware feedback from the analytical quantum cost model. The process jointly optimizes accuracy, total cost, and parameters, producing Pareto-optimal HQNN architectures with hardware-aware cost estimation.}
    \label{fig:methodology}
    \vspace{-13pt}
\end{figure*}

\vspace{-8pt}
\section{Proposed Hyb-HANAS Framework}\label{sec:HANAS_framework}
\vspace{-2pt}

We integrate the proposed analytical quantum and classical cost models into a  NAS framework, introducing Hyb-HANAS, which finds HQNN architectures optimized for accuracy, parameter count, and computational cost. The HQNN architecture used is similar to Fig.~\ref{fig:F_and_t_vs_Q_andD}-top.
Hyb-HANAS supports two search configurations: (1)~a quantum-focused search, where classical layers are fixed and only the quantum layer design is optimized, useful for benchmarking QNNs against classical networks, and (2)~a full hybrid search, jointly exploring both classical and quantum components. An overview of 
the proposed framework is shown in Fig. \ref{fig:methodology}.
%
\vspace{-8pt}
\subsection{Phase1: Initialization and Configuration}
\vspace{-2pt}

\subsubsection{\textbf{Classical Mode Selection and Search Space Configuration}}

Hyb-HANAS begins with an interactive setup where the user is prompted to configure the classical CNN component. Two modes are supported: \textit{fixed} and \textit{variable}. 
In fixed CNN mode, the user is required to enter the number of CNN layers (between 1-3) along with their hyperparameters (output channels, kernel sizes, stride values, pooling, and dropout). Once defined, they remain fixed throughout the search, and only the quantum layer is updated. For variable CNN mode, 
the user needs to add a set of values for each hyperparameter. The evolutionary search then explores diverse CNN architectures, alongside the quantum layer, within the user-defined bounds.
The framework, by default, utilizes angle embedding for data encoding. Since angle embedding requires one qubit per feature \cite{lloyd2020}, to ensure consistent interfacing between classical and quantum components, a projection layer is added to automatically map CNN outputs to the quantum input dimension. A final classical linear layer is also added, whose neurons are equal to the number of classes in the dataset (e.g., $10$).

In our experiments, we explore both \textit{fixed} and \textit{variable} CNN modes.
In the fixed mode, a single convolutional layer with $16$ output channels, kernel size $k=3$, MaxPool (stride $s=2$), dropout rate $0.1$, and \texttt{ReLU} activation is used.
In the variable mode, up to three convolutional layers are explored, with output channels ${8,16,32,64}$, kernel sizes ${3,5,7}$, activations \{ReLU, SiLU, Tanh\}, an optional average pooling, and dropout (rate = 0.1 when applied).
\vspace{-5pt}
\subsubsection{\textbf{Quantum Backend Selection and Search Space Configuration}}

After classical mode and search space configuration, users then select the quantum backend, either real IBM Quantum device or simulated fake devices \cite{IBM_fake_backends}. For real backends, the framework requires authentication via the IBM Quantum API. 
Fake backends emulate hardware characteristics such as coupling maps, basis gates, and qubit properties ($T_1$, $T_2$, gate errors). These simulated backends enable realistic, noise-aware benchmarking and quantum cost estimation without premium access to real IBM devices; however, they can often be outdated. 
For the quantum component, users define the search ranges for quantum parameters such as qubit count, depth, entanglement gates, and gate types. In our experiments, qubit counts span \{2-12\}, parameterized rotation gates can be \{RX, RY, RZ\}, and entanglement topologies include \{linear, circular, full, star, grid\}~\cite{mordacci2025impact}. Two-qubit gates are selected from \{CNOT, CZ\}, and circuit depth varies from \{1-15\} layers, enabling exploration of trade-offs between expressibility, noise resilience, and trainability. 
%
%
%
\vspace{-4pt}
\subsubsection{\textbf{Backend Profiling Module}}
Real-time calibration data is retrieved from the selected IBM Quantum backend to support the analytical cost model (Section~\ref{sec:Qcost_model}). Accessing this information in real time significantly minimizes the impact of calibration drift, as IBM devices are typically recalibrated 
daily~\cite{IBM_calibration}. By aligning the profiling step with the most recent hardware state, this module ensures that the estimated quantum cost reflects the backend’s current operational characteristics.
It collects per-qubit coherence times ($T_1$, $T_2$), readout durations and errors, average gate durations, and error rates for single- and two-qubit operations. All data is cached using Python’s \texttt{@lru\_cache()} decorator to avoid repeated API calls and automatically refresh upon backend recalibration. This ensures that cost estimation remains hardware-accurate while minimizing query overhead during NAS.
\vspace{-7pt}
\subsubsection{\textbf{Dataset}}\label{subsec:dataset}
Once the configuration and search space setup are complete, the data pipeline module prepares the MNIST dataset for evaluation. We use all ten classes, selecting 400 samples per class (4,000 total), with 80\% training and 20\% validation split.
This reduced dataset is a standard practice in hybrid NAS, as it significantly decreases the computational load of evaluating hundreds of HQNN models, many of which include computationally intensive quantum circuit simulations, while still preserving reliable architecture ranking.




\vspace{-8pt}
\subsection{Phase 2: NSGA-II Multi-Objective Search}
\vspace{-2pt}

The 
hardware-aware hybrid architecture search 
jointly optimizes all the objectives, i.e., $\min \big( 1 - \text{Acc}, T_{\text{quantum}}, T_{\text{classical}}, \text{Params}_{\text{total}} \big)$, where $T_{\text{quantum}}$ and $T_{\text{classical}}$ represent the analytically estimated 
time, for the quantum and classical parts, respectively, and $\text{Params}_{\text{total}}$ is the total parameters in the hybrid network.

\textbf{\textit{Population initialization.}}  
We consider a total of 8 generations, each with a population size of 12. While the search space can be expanded, our goal in this paper is not to find universally-optimal HQNNs, but to demonstrate the effectiveness of hardware-informed multi-objective optimization in HQNNs, based on our analytical model. At generation $G=0$, the initial population of HQNNs is randomly sampled from the defined search space. Each gene encodes both quantum (qubit count, gate types, entanglement topology, and circuit depth) and, if applicable, classical CNN parameters (layers, kernels, pooling, activations). When the CNN is fixed, only quantum parameters are encoded.

\textbf{\textit{Evolutionary loop.}}    
For $G>0$, new offsprings are generated 
using genetic operators, i.e., \textit{crossover}, which combines parent attributes to form mixed HQNN designs
, and \textit{mutation}, which introduces small random variations in qubit count, gate type, kernel, etc. 
These operators are applied to every offspring, with 
each architectural gene independently perturbed with probability $p_m=0.4$ to maintain diversity in the search population.

\textbf{\textit{Evaluation and objective computation.}} 
Each candidate architecture (CNN + quantum layer + classifier) is trained for $10$ epochs using the Adam optimizer (lr = 0.01). 
Following the standard NAS practices, an early stopping technique (stop further training, if val acc $<$ 20\% after two epochs), is used to evaluate a large number of candidate HQNN architectures efficiently while maintaining meaningful architectural trends.
After training, three metrics are recorded: validation accuracy, total hardware cost ($T_{\text{quantum}}+T_{\text{classical}}$), and total parameter count.
For classical cost estimation (section \ref{sec:classical_cost_model}), the quantum layer is excluded, so that the measured FLOPs and latency represent only the classical computation. Including quantum layers would artificially inflate the classical cost, since quantum layers, although executed on an external quantum backend, still introduce additional latency during forward and backward passes despite not contributing to classical arithmetic operations.

For efficiency, Hyb-HANAS operates as a \emph{hardware-in-the-loop emulation} framework, i.e., quantum layers are trained on the simulator using efficient gradient methods (i.e., adjoint differentiation~\cite{adjoint_diff}), while their hardware resource cost is analytically estimated from real backend calibration data. This design enables scalable exploration without direct hardware runs, which can be significantly expensive and slow due to limited shot counts, queuing delays, and restricted access to NISQ devices. 

\textbf{\textit{Selection mechanism.}}  
A fast non-dominated sorting algorithm ranks architectures into Pareto fronts based on objective dominance. In each front, the crowding-distance metric preserves diversity by favoring sparsely populated regions. Parents are selected via rank-based tournaments, and offsprings are produced through crossover and mutation. The next generation is formed by merging parents and offspring while retaining the most diverse non-dominated individuals until the population size is restored.
\vspace{-9pt}
\subsection{Phase 2A: Analytical Quantum Cost Model}
\vspace{-2pt}

The hybrid models are implemented in \texttt{PennyLane}, which supports seamless quantum-classical integration~\cite{qlayers_pytorch}. 
To estimate the quantum layer's cost, for each candidate hybrid architecture, quantum layers are converted into IBM QASM circuits, transpiled to the selected backend, and the true hardware cost is evaluated using real calibration data, as explained in Section \ref{sec:Qcost_model}. 
\vspace{-8pt}
\subsection{Phase3: Post-search Analysis}
\vspace{-2pt}

Upon completion of the NSGA-II search, the resulting non-dominated population forms a Pareto front representing optimal trade-offs among accuracy, total hardware cost, and parameter count.
The 2D Pareto plots illustrate pairwise trade-offs, such as accuracy versus quantum time or parameter count, highlighting how different architectures balance performance and hardware efficiency along individual objectives.
The 3D plots visualize the full multi-objective landscape across accuracy, total cost, and parameter count, showing the jointly optimized architectures.
Additionally, a detailed breakdown of the quantum cost components (e.g., routing overheads and reliability penalties) quantifies their relative contribution to the total quantum cost.
%
\vspace{-9pt}
\section{Results and Discussion}\label{sec:results}
\vspace{-3pt}

\subsection{Analytical Quantum Cost Validation}\label{subsec:qiskit_scheduler_comparison}
\vspace{-2pt}

Direct runtime comparison with IBM Quantum backends is not meaningful, as reported times include queuing, control, and I/O overheads, which are not related to actual quantum hardware cost. However, we validate our analytical model against IBM Qiskit’s built-in scheduler~\cite{qiskit_scheduler}, which is used to estimate circuit execution windows from calibrated pulse durations. This corresponds to the total gate time in the analytical model (Step 1 in Algorithm \ref{alg:qcost_model}), as it does not include noise-induced penalties.
\begin{figure}[t!]
    \centering
    \includegraphics[width=0.99\linewidth]{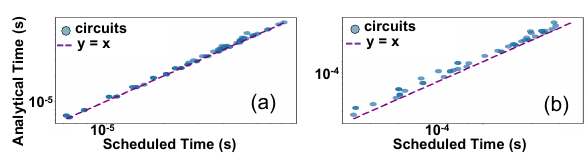}
    \vspace{-4pt}
    \caption{Analytical model's estimated time vs. IBM Qiskit scheduler time one (a) \texttt{ibm\_torino} and (b) \texttt{FakeWashington}. }
    \label{fig:analytical_vs_device}
    \vspace{-16pt}
\end{figure}
%
%
Fig. \ref{fig:analytical_vs_device} shows that execution times estimated by our analytical model and Qiskit's scheduler on two different backends align closely across $100$ random circuits (2-5 qubits, depths 50-600). However, the time estimated by our analytical model is slightly higher than the scheduler's estimated time, because Qiskit treats certain rotations as zero-time~\cite{qiskitrzdoc}, whereas the analytical formulation accounts for all calibrated physical gates. The close alignment in cost estimation validates that the proposed model accurately reproduces hardware cost.


\vspace{-8pt}
\subsection{Hyb-HANAS: Fixed CNN Mode Results} 
\vspace{-2pt}

\begin{figure*}[h]
    \centering
    \includegraphics[width=0.99\linewidth]{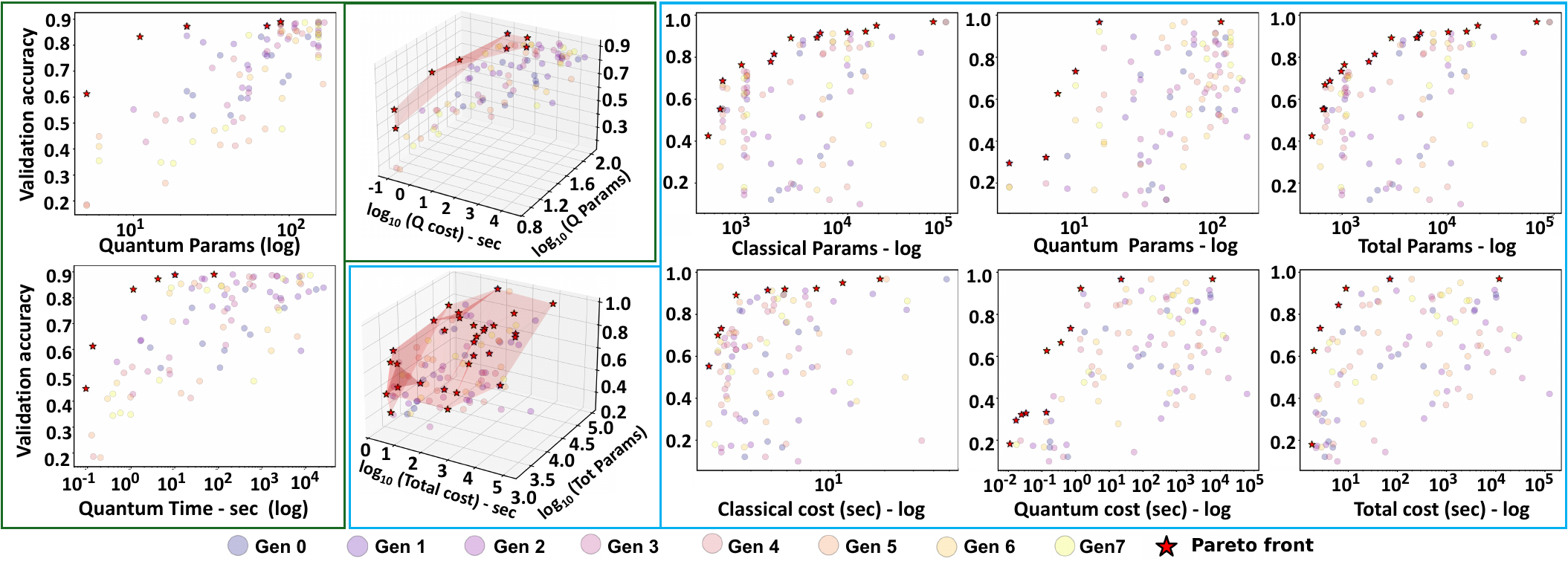}
    \vspace{-4pt}
    \caption{Pareto-optimal architectures obtained under fixed (enclosed in green) and variable (enclosed in blue) CNN mode. In fixed CNN mode, NSGA-II optimizes quantum design parameters while the classical CNN remains constant. The 2D plots illustrate trade-offs between two objectives, namely accuracy vs. quantum cost, and accuracy vs. quantum parameter count. In variable CNN mode, both classical and quantum layers are evolved during search, resulting in a richer Pareto front. The 2D plots highlight trade-offs between individual objectives w.r.t accuracy. The 3D plots in both settings show Pareto optimal architectures in all objectives. Color-coded generations (purple-to-yellow) show evolutionary convergence toward the Pareto front (red stars).}
    \label{fig:results_both_fixed_variable}
    \vspace{-12pt}
\end{figure*}

The fixed CNN mode isolates the quantum layer’s influence on performance by keeping the classical part constant, enabling focused analysis of quantum design sensitivity, i.e., how qubit count, circuit depth, etc, affect accuracy and hardware cost. This setting is also valuable for benchmarking hardware-aware quantum optimization, focusing only on evaluating the efficiency and expressiveness of quantum layers.

In this setting, NSGA-II explores only the quantum design space, optimizing trade-offs between validation accuracy, quantum parameters, and hardware cost. Since the classical part is fixed, all candidate architectures share the same classical metrics and collapse into a vertical line, resulting in an uninformative Pareto front (as discussed in Section \ref{subsec:mot_analysis}). Therefore, we only report Pareto-optimal architectures w.r.t. quantum parameters and quantum cost, as shown in Fig.~\ref{fig:results_both_fixed_variable} (enclosed in green). 
The 2D Pareto plots illustrate validation accuracy versus quantum time and quantum parameter count, showing that efficient, shallow circuits can achieve comparable accuracy to deeper or more entangled ones that incur higher hardware cost. 
The 3D plot visualizes Pareto-optimal architectures that jointly balance quantum parameters, quantum hardware cost, and accuracy, identifying hardware-efficient yet expressive quantum designs. These results demonstrate that even with a fixed classical component, accuracy-cost diversity arises purely from quantum structural variations, offering valuable insight into the role of quantum circuit parameters in hybrid model optimization.
%
%
%
\vspace{-8pt}
\subsection{Hyb-HANAS: Variable CNN Mode Results}
\vspace{-2pt}

In this configuration, both the classical and quantum components are optimized jointly by Hyb-HANAS, allowing co-adaptation across design spaces. Unlike the fixed-CNN mode, the search evolves through CNN depth, channels, kernels, pooling, etc, alongside quantum hyperparameters such as qubit count, depth, and entanglement topology, highlighting the co-adaptation between classical and quantum processing. 
The results are presented in Fig.~\ref{fig:results_both_fixed_variable} (enclosed in blue). The 2D Pareto plots highlights trade-offs between accuracy and major cost factors, i.e., classical, quantum, total combined cost, and total parameters, while the 3D Pareto surface shows the architectures that jointly balance all the objectives. Compared to the fixed CNN case, the inclusion of the classical part in the search results in a richer diversity of trade-offs, where both CNN and quantum layer's complexity jointly influence the objectives. These results demonstrate that multi-objective co-optimization across classical and quantum components is essential to \mbox{identify} \mbox{globally} efficient HQNNs that exploit the complementary strengths of both domains.
\vspace{-8pt}
\subsection{Quantum Cost Ablation Analysis}\label{sec:ablation_study}
\vspace{-2pt}

\begin{figure}[t!]
    \centering
    \includegraphics[width=1.0\linewidth]{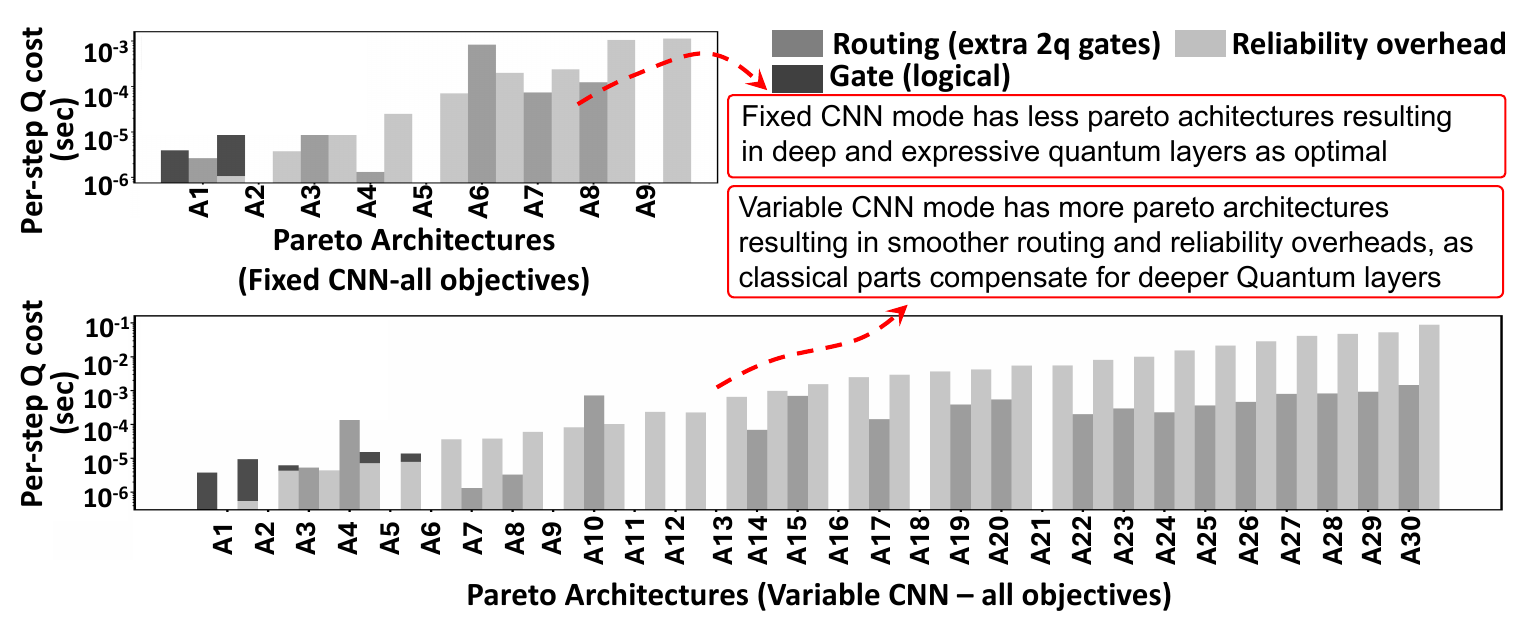}
    \vspace{-4pt}
    \caption{Quantum cost ablation analysis for Pareto-optimal architectures in fixed (top) and variable (bottom) CNN modes. Bars show per-step quantum time decomposed into logical gate, routing, and reliability/sampling penalty components. Fixed CNN mode highlights higher routing and reliability overheads for deeper circuits, while variable CNN mode yields smoother, hardware-efficient cost distribution through classical-quantum co-adaptation.}
    \label{fig:contrib_plots}
    \vspace{-12pt}
\end{figure}

To analyze how hardware-level effects contribute to the total quantum hardware cost, we decompose the per-step quantum cost of global Pareto-optimal architectures into (1) logical gate time, (2) routing overhead from additional two-qubit gates, and (3) reliability penalties arising from gate errors and decoherence. The results of the ablation analysis are presented in Fig.~\ref{fig:contrib_plots}.
In the fixed-CNN mode, the search explores only the quantum design space, resulting in a few Pareto-optimal architectures. Most of these exhibit large routing and reliability penalties, indicating the use of deeper quantum layers, and complex entanglement patterns, which eventually force the transpiler to insert extra two-qubit gates to meet connectivity constraints and eventually greater error accumulation. 
On the other hand, the variable-CNN mode yields a larger number of Pareto-optimal architectures, with routing and reliability contributions ranging from very small to relatively large. This richer spectrum highlights the co-adaptation of classical and quantum parts during the search. As classical CNN depth or feature extraction capacity increases, the Hyb-HANAS can select simpler, shallower, and better-routed quantum circuits, resulting in reduced hardware penalties while maintaining accuracy. Similarly, when the classical CNN is lightweight, the search compensates by adopting more expressive quantum designs, even when they incur higher hardware cost.

As an ablation study, we report the per-step quantum cost rather than the total training cost to provide a normalized hardware-specific breakdown that is independent of dataset size or training epochs. We do not show gradient-evaluation overhead from the parameter-shift rule (Step 5, Algorithm \ref{alg:qcost_model}) because it scales uniformly across all architectures. 
Overall, the ablation results show that routing and reliability penalties dominate as circuit complexity increases, highlighting the need for compact, well-routed, and noise-resilient quantum layers in HQNNs.

\vspace{-8pt}
\section{Conclusion}\label{sec:conclusion}
\vspace{-2pt}

We proposed an analytical quantum cost model and a complementary classical cost model to provide a unified, cross-domain metric which expresses both quantum and classical computational resource requirement (in HQNNs) in the same physical unit, i.e., seconds. This unified formulation enables single yet interpretable metric for hardware-resource estimation in HQNNs and in general hybrid algorithms. By incorporating these models within an NSGA-II–based neural architecture search (NAS) framework, we proposed Hyb-HANAS, which jointly optimizes accuracy, network parameters and hardware cost of both classical and quantum components in HQNNs under realistic hardware constraints.
The analytical quantum model is built from backend calibration data in real-time and accounts for gate durations, routing overheads, and noise-induced sampling inefficiencies. Our method enables hardware-resource estimation of quantum layers in HQNNs without requiring to execute on real devices. The proposed quantum cost model not only guides efficient hardware-aware HQNN design, it can also be extended to other scenarios including, but not limited to quantum circuit compilation benchmarking, hardware selection and scheduling, algorithmic fidelity-latency trade-off analysis, and training-time estimation for near-term NISQ devices.
\section*{Acknowledgements}
\vspace{-0.14cm}
This work was supported in part by the NYUAD Center for Quantum and
Topological Systems (CQTS), funded by Tamkeen under the NYUAD Research
Institute grant CG008.
\bibliographystyle{ACM-Reference-Format}
\bibliography{main}

\end{spacing}
\end{document}